# The vimentin cytoskeleton: When polymer physics meets cell biology

Alison E. Patteson[1,2], Robert J. Carroll[1,2], Daniel V. Iwamoto[3], Paul A. Janmey[3,4]

1 Physics Department, Syracuse University, Syracuse, NY 13244
2 BioInspired Institute, Syracuse University, Syracuse, NY 13244
3 Institute for Medicine and Engineering, Department of Physiology, University of Pennsylvania, Philadelphia, PA 19104
4 Department of Physics and Astronomy, University of Pennsylvania, Philadelphia, PA 19104

**Abstract**
The proper functions of tissues depend on the ability of cells to withstand stress and maintain shape. Central to this process is the cytoskeleton, comprised of three polymeric networks: F-actin, microtubules, and intermediate filaments. Intermediate filament proteins are among the most abundant cytoskeletal proteins in cells; yet they remain one of the least understood. Their structure and function deviate from those of their cytoskeletal partners, F-actin and microtubules. Intermediate filament networks show a unique combination of extensibility, flexibility and toughness that confers mechanical resilience to the cell. Vimentin is an intermediate filament protein expressed in mesenchymal cells. This review highlights exciting new results on the physical biology of vimentin intermediate filaments and their role in allowing whole cells and tissues to cope with stress.

**1. Introduction**

Mammalian cells rely on the cytoskeleton, a dynamic system of polymer networks, to generate changes in cell shape while preserving mechanical strength. The cytoskeletal network of animal cells is comprised of three biopolymer networks: F-actin, microtubules, and intermediate filaments (IF). Unlike F-actin and microtubules, which are highly conserved and expressed in nearly every eukaryotic cell type, there are many diverse forms of IFs, and they are expressed in tissues in a cell type-specific manner. Intermediate filaments are expressed mainly in animals [6], their origin corresponding to the evolution of multi-cellularity and their associated physical challenges. There are a number of discerning and enigmatic ways in which IF differ from polymers of actin (F-actin) and tubulin (microtubules). IF polymerize without nucleotide hydrolysis and form filaments with no polarity. There are no known motor proteins that move along IFs, although IF can be carried as cargo by motors that move along microtubules or F-actin. More so than actin and microtubules, intermediate filaments serve as cytoskeletal links that transfer forces between the cell surface and nuclear surface, which makes them a prime candidate for generating new insight into how individual cells and whole tissues cope with various types of stress.

Vimentin is classified as a Type III IF, meaning that vimentin IFs are formed by polymerization of a single protein, unlike most other IFs, such as keratin filaments or neurofilaments, which are copolymers of multiple distinct, but related gene products. In vertebrates, vimentin is expressed primarily in mesenchymal cells as found in connective and adipose tissues, though it is also expressed in early development, or as a result of wounding, malignant transformation of epithelial cells, or other disease-related settings. IFs in general and vimentin IF in particular have many different functions including mediating signal transduction, sequestration or activation of proteins and nucleic acids, and acting as cell surface cofactors for attachment of some pathogens [8, 9], but a critical aspect of their cellular function appears to be mechanical. The mechanical function of IFs is evident in the consequence of mutations in the genes that code for them, which can lead to human diseases characterized by skin blistering and other clear mechanical defects [10]. This review focuses on the mechanical features of vimentin filaments and the networks they form, both *in vitro* and *in vivo*, with an emphasis on how ideas and methods in soft matter physics have been effective in illuminating the biological function of this protein.

Among the three cytoskeletal filaments, IFs stand out by being the most flexible. IFs self-assemble into linear polymers with a diameter of approximately 10 nm and lengths of several microns [11]. Before single molecule pulling experiments were possible, the bending stiffness of these filaments was calculated from their persistence length $l_p$: the length scale for the decay of the tangent-tangent correlation along the filament. Assuming that there is no intrinsic curvature in the structure of the filament and that only thermal energy distorts its contour, the persistence length $L_p$ is related to the bending stiffness, $\kappa_b$, of the filament by the relation $L_p = \kappa_b / k_B T$. If a polymer can be modeled as a uniform cylinder, then the Young's modulus E can be calculated from the bending stiffness, which is scale-dependent, by the relation $E = \kappa_b / I$, where $I = \pi d^4 / 64$ is the area moment of inertia and d is the filament diameter. The persistence length of a filament can be measured by imaging it adsorbed to a surface, tracing its contour and then calculating tangent-tangent correlations, after taking into account distortions of the filaments as they adsorb from three-dimensional (3D) thermally writhing polymers onto an adhesive surface [12]. Such measurements have led to a consistent hierarchy of stiffness ranging from $L_p$ > 1 mM for microtubules to 10 µm for F-actin to 1 µm for vimentin filaments [13]. The Young's modulus calculated from the persistence length of 1 µm for vimentin and assuming a diameter of 10 nm is 8 MPa.

## 2. Mechanical response of vimentin networks and filaments *in vitro*

On a macroscopic scale, the most striking property of IF networks is their stiffening at increasing strains, and their ability to withstand large strains and resist large stresses that would otherwise rupture actin and microtubule networks [14] (Figure 1). This unusual resistance to breakage and ability to deform to very large strains is also observed in single filament studies [2, 15]. In one study a single vimentin filament was adsorbed onto an adhesive surface across a 250 nm diameter hole, providing a free portion of the filament on which a 3-point bending test was performed using an atomic force microscope with a sharp tip that indented the middle of the filament. This enabled calculation of the bending modulus from the relation between force and indentation depth. Moreover, these studies showed that the response of the filament to indentation is a combination of stretching, quantified by a Young's modulus, and shearing, due to sliding of protofibers within the filament. As the filament elongates, it also thins, and so the bending modulus decreases as the filament is deformed. When the indentation is done relatively fast (in the range of 1 Hz) the response is primarily elastic. These measurements show that the Young's modulus of a vimentin filament is between 300 and 400 MPa, which is much lower than that of an actin filament (2 GPa) [16] or a microtubule (1.2 GPa) [17]. The estimate of E from stretching the vimentin filaments is much larger than the value of E of 8 MPa estimated from the persistence length, suggesting that vimentin filaments might be much more flexible to small amplitude bending motions than to longitudinal stretching. When the vimentin filament was crosslinked by glutaraldehyde, the bending modulus increased to 900 MPa, consistent with the expectation that glutaraldehyde fixation prevents subunit sliding, but this value is still much smaller than that of F-actin or microtubules. Single vimentin filament stiffness has also been measured by optical trapping studies in which two beads are attached to opposite ends of a filament and then one bead is pulled away from the other, with the resulting force detected by displacement of the bead within the optical trap. Figure 2 and the inset to this figure show that the force rises linearly with elongation up to a strain of approximately 10%, after which there is first a decrease in slope, suggesting a softening, and then an increase, indicating stiffening at strains above 50%. Use of an AFM that pulls the filament at a constant speed can extend the strains to which a single vimentin filament can be pulled. These data, which agree with optical trapping, as well as with the 3 point bending results, show that the stiffening response continues to strains over 300% [2]. A force of about 8 nN was required to break a single vimentin filament, which contrasts to the much smaller force of 300-600 pN that is sufficient to break an actin filament [16], even though the stiffness of F-actin is much higher.

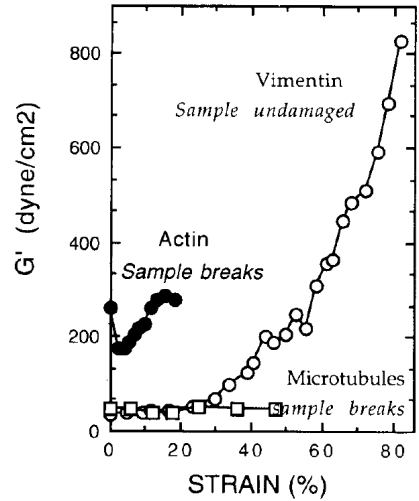

Figure 1. Nonlinear elasticity of vimentin networks. The shear modulus of vimentin gels increases with increasing strain and the sample remains intact at large strains where actin and microtubule networks fail. Adapted from [1].

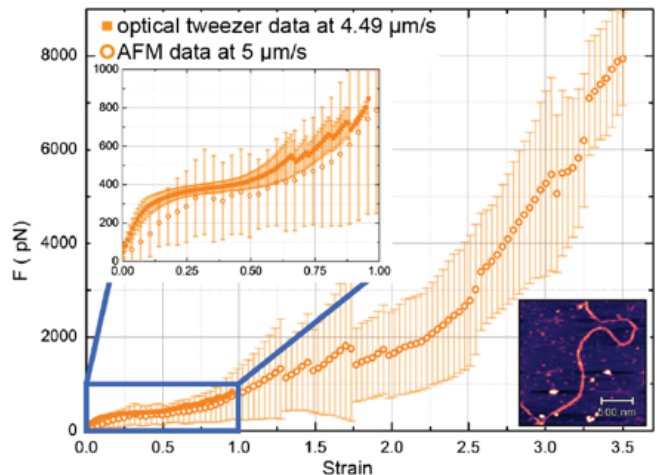

Figure 2. Force response to increasing uniaxial strain of vimentin filament measured by AFM (open circles) or optical trap (closed squares) Bottom right inset: AFM image of a vimentin filament. From [2].

## 2.1 Molecular structure of vimentin filaments

The remarkable mechanical properties of vimentin filaments are explained by the molecular structures of the vimentin filament subunits and in the way these are packed together in the filaments (Figure 3). Unlike actin and tubulin, for which there exist atomic or near atomic level resolution structures, and in which most of the protein is folded, vimentin and other IF proteins have resisted crystallization, except for small fragments, and cryo-EM imaging has not yet lead to atomic resolution models of their structure. The amino acid sequence of IF proteins is predicted to be largely alpha helical and has the appropriate spacing of specific amino acids required to make an alpha-helical coiled coil of two monomer units in parallel.

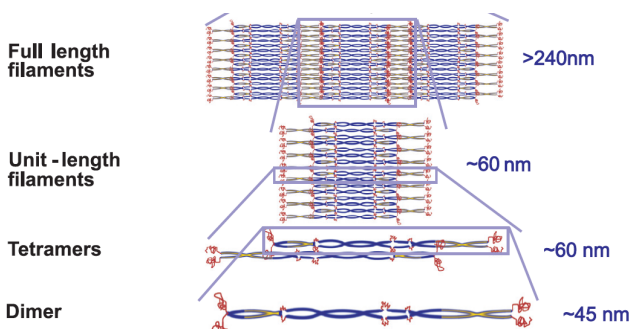

Figure 3. Schematic of the steps in vimentin filament assembly. Adapted from [3].

Most of the protein interior is predicted to be alpha helical, but there are structurally important interruptions to this pattern that break the helix, and both the N- and C-termini appear to be unstructured and likely to extend away from the surface of the coiled coil when it assembles into a filament. When the polypeptides emerge from the ribosome, they are immediately woven into 45 nm long coiled coil dimers, but dimers appear not to be stable in the cell unless they also bind another coiled coil dimer to form a tetramer. The two coiled coil dimers bind in an antiparallel manner with a partial overlap. As a result, this stable tetramer unit, the smallest subunit from which filaments assemble, is a symmetrical rod-like object with a length of 60 nm and a width of a few nm.  At very low ionic strength and moderate urea concentrations, tetramers can be stable *in vitro*, but under more physiologically realistic conditions tetramers rapidly assemble by lateral interactions into complexes that contain approximately 8 tetramers and are slightly longer than 65 nm [2, 18].  These unit length filaments (ULF) then assemble end-to-end by annealing the tetramers that extend from the ends of the ULF to form the vimentin intermediate filament.  At this point, at least *in vitro*, the polymerization reaction is not yet complete. In a process that is slower than the assembly of ULFs into a filament, the diameter of the vimentin filament decreases from approximately 15 nm to 10 nm in the mature filament [19]. Some compaction also occurs at the level of the ULF formation as tetramers assemble stepwise by lateral binding [20].  Since the filament diameter decrease occurs without a change in the mass/length ratio and no energy consumption is needed [19], water is being removed from the filament, presumably as polypeptide chains reorient to form additional bonds and decrease their free energy.

The unusual stability of IFs against mechanical rupture depends on the formation of many simultaneous bonds along the filament diameter, many of which are electrostatic or hydrogen bonds.  Unlike actin nor tubulin polymers, in which the binding interfaces between subunits are largely hydrophobic, the bonds between dimers in the tetramer and among tetramers in the filament involve significant overlap between local regions of opposite charge.  These interactions involve not only regions within the alpha helices, but also between the N-terminus and the core of the filament [21]. The vimentin N-terminus is especially important, in part because it is among the few domains with a large net positive charge in this overall highly negatively charged protein and the strong polyelectrolyte filament that it forms.  The N-terminus of vimentin is targeted by several protein kinases that phosphorylate neutral serine residues to bestow a negative charge, and protein arginine deiminases that convert positively-charged arginine to neutral citrulline [9].  These alterations of charge can lead to filament disassembly, and at lower extents can decrease the stiffness of the filament [22].

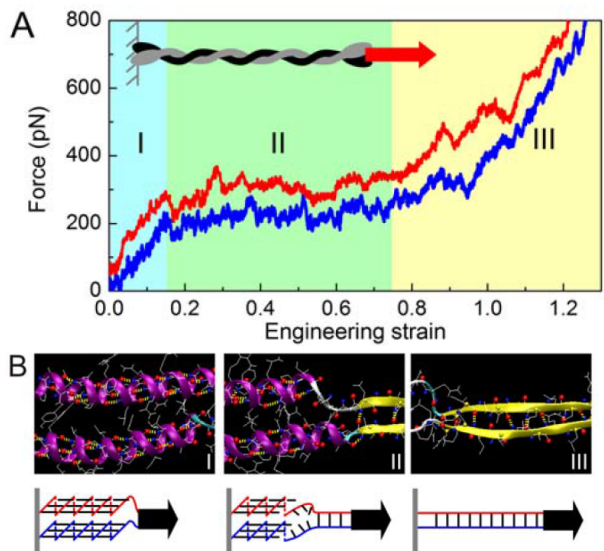

Figure 4. Stiffening and unfolding of a vimentin dimer under tension. A: force–strain relations for tensile deformation of the vimentin dimer at two pulling speeds. B: snapshots from molecular dynamics analysis of a region of vimentin that unfolds in the three ranges of strain shown in A. Adapted from [5].

## 2.2 Relating vimentin structure to mechanical response

Since the fundamental covalent unit of the vimentin filament is the coil-coil dimer, the mechanical properties of this structure, especially in elongation, are central to the mechanical response of the filament. Individual vimentin dimers have also been pulled by atomic force microscopy, and the response to elongation is strikingly similar to that of the whole filament. Figure 4A shows how the force increases when the coiled coil dimer is stretched up to 120% for two different rates of extension. Molecular dynamics simulations (Figure 4B) suggest that this large strain can be accommodated by converting alpha helical structures to beta strands, where amino acids from both monomers still remain hydrogen-bonded, but adapt to a large increase in contour length [3]. The Young's modulus derived from stretching a single coiled coil is approximately 400 MPa, which is very similar to that of the whole filament, suggesting that at least at small strains, the filament responds to stretching by distributing the force over all the coiled coils in parallel.

## 2.3 Inelastic responses and structural reorganization of vimentin filaments and networks

When the deformations of the vimentin filament are relatively small the response of the filament to mechanical stress is largely elastic, and the filament returns to its initial state without dissipating energy. However, at larger strains, structural changes that dissipate energy occur, as for example, the alpha helix transforms to a beta strand, or the filament thins as subunits slide past one another. These viscous or plastic deformations are made possible without damage to the filament because the bonds that hold the dimers in the tetramers and the tetramers within the filaments are numerous and individually weak, and are, compared to high affinity enzyme-substrate binding, relatively nonspecific. The structural transitions that occur when vimentin filaments are stretched and the degree to which they recover when the deforming force is released have been studied in increasing detail. Figure 5 shows the relation between force and strain as a single filament is stretched at a constant rate and then returns to its resting length for multiple rounds of cyclic loading at increasingly large strains [4]. These results show that the larger the strain, the greater is the energy dissipated during the deformation of the filament, a metric that is related to the area between the loading and unloading curves. Also, each successive round of elongation causes the initial slope to decrease, meaning that the filament becomes softer after each elongation cycle. Even when the filament strain exceeds the linear elastic response range, the recovery of its initial state after the force is removed can be complete, as the conformational changes are reversed. Even at large deformations the force is finite until the strain is returned to zero, indicating that some aspect of the filament structure is not destroyed by the elongation cycles. The energy dissipation is related to the structural transitions within the alpha-helical coiled coils, but other rearrangements are also possible. When the strain exceeds a range at which substantial helix to strand changes occur, the recovery is incomplete and the filament softens after repeated deformation and transitions into a new resting state [23].

In many respects, the viscoelasticity of vimentin networks reflects the properties of individual filaments. Interpreting the rheology of vimentin networks is complicated by the fact that there are no known crosslinkers of vimentin or other IFs analogous to filamin A or alpha-actinin which are well characterized crosslinkers of F-actin. Therefore the network geometry, and more importantly the degree of connectivity at vimentin crosslink points, is unknown. Most rheological studies of vimentin have used divalent cations as linkers between filaments or relied on steric interaction and weak inter-filament bonds to stabilize the network and provide it an elastic response. However, consistent with the much lower bending stiffness of vimentin filaments compared to F-actin, the elastic modulus of vimentin networks is lower than that of a comparable concentration of actin, and vimentin gels can withstand much larger strains without breaking [1, 24-26]. Most studies of vimentin networks have measured shear deformations, but threads of vimentin can also be formed, and these, like individual filaments, resist stretching to very large strains without rupture [27]. However, when vimentin networks are deformed to sufficiently large strains they also undergo softening and fluidization, but unlike individual filaments where these effects are related to transitions in protein secondary structure, in networks the fluidization is due to loss of inter-filament attachments [28].

An open question about the mechanism of elasticity in vimentin networks is whether theoretical models designed to explain the strain-stiffening and compression-softening effects in actin networks and other crosslinked semiflexible polymer networks [29-33] also apply to vimentin. These models are appropriate for systems in which the filaments are sparsely connected with coordination at junction sites between 3 and 4, and where the persistence length of the filaments is larger than the average mesh size of the network. Since the mesh size of vimentin networks at experimentally attainable concentrations is on the order of 100s of nms to 1 µm, and the persistence length of vimentin is $\leqslant$ 1 µm, the assumptions of these models are not obviously fulfilled, and the unusual mechanical responses of vimentin might require additional theoretical work to understand.

## 3. Assembly into larger vimentin networks and bundles

### 3.1. Counterion-mediated bundling effects

The vimentin filament is a strong polyelectrolyte. Its net negative surface charge density is -0.5/nm$^2$, which is more than twice as great as that of F-actin and almost as large as DNA [34]. The distance between fixed charges on the filament surface is much smaller than the Bjerrum length, the distance at which electrostatic energy is equal to k$_B$T, and therefore positively charged counterions will condense on its surface. If the counterions are multivalent, their condensation results in attractive interactions between filaments that lead to filament crosslinking into networks [26, 35] or bundles [36]. In the cell, various counterions can affect the structure of vimentin filaments [34]. These cations alter the morphology of the vimentin network, resulting in changes in the mesh size and stiffness that are dependent on ion concentration and valency. Ion pumps along the cell membrane control counterion movement into and out of the cell, and physiologically relevant divalent cations such as $Mg^{2+}$, $Ca^{2+}$, $Cu^{2+}$ and $Zn^{2+}$ have much larger effects than monovalent ions. Counterion effects are usually studied in reconstituted networks of purified intermediate filaments, where the bulk mechanical properties can be more directly measured. At low concentrations, divalent cations behave as effective crosslinkers in the network [26, 37], increasing the network stiffness in a manner consistent with the theory of semiflexible polymers [38]. In addition to the general attraction of counterions to the filament surface, crosslinking is also mediated by the C-terminal tail domain of vimentin [26]. At higher concentrations, the vimentin filaments begin to form lateral filament-filament structures called bundles. At sufficiently high cation concentration, the bundling can actually lower the stiffness of the network due to the loss of percolation [39], if the interaction strength between filaments in the bundles strong enough to disrupt elastically active crosslinks [40].

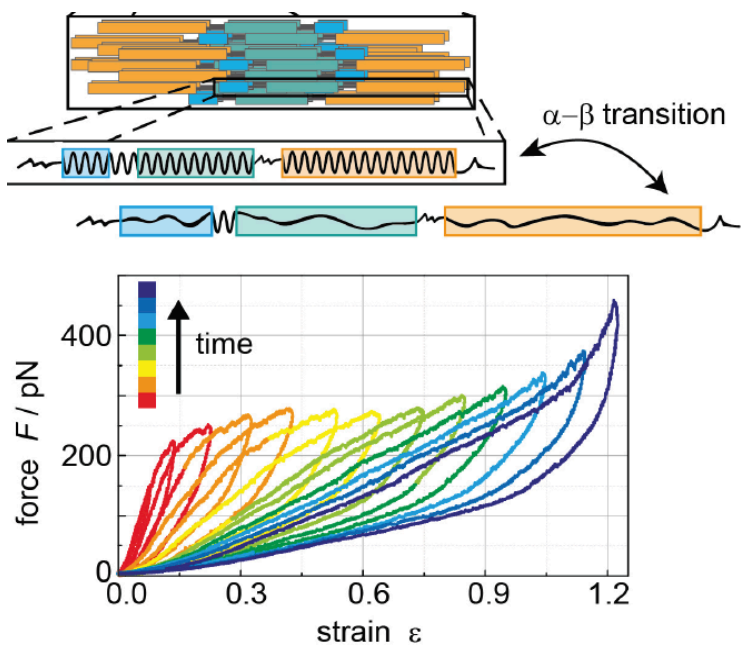

Figure 5. Force during cycles of increasing uniaxial strain followed by return to initial length for a series of consecutive deformation circles at increasing maximal strains at a constant rate of ~1 μm/s. The diagram above shows the increase in length that results from a transition from alpha helix to beta strand. Adapted from [4]

The morphology of the vimentin networks depends on the ion valency; reconstituted vimentin networks containing divalent cations with valence of 2+ form thicker filaments than monovalent cations with valence 1+ [41]. Reconstituted vimentin networks also bundle more readily with divalent cations [42, 43]. A microfluidic study showed that bundling occurs at multivalent cation concentrations of around 50% of the counterions *in vitro* [44]. The counterion species also makes a difference; bundling can occur for *in vitro* vimentin networks at concentrations of 10 mM for $Ca^{2+}$ and $Mg^{2+}$ but at lower concentrations for $Mn^{2+}$ ions [34]. A recent *in vitro* study [39] showed that bundle formation occurs for $Zn^{2+}$ at a concentration as low as 100 μM, and the effect of $Zn^{2+}$ in facilitating the lateral interactions between vimentin filaments necessary for bundling is mediated not only by electrostatic interactions but also by the ability of $Zn^{2+}$ to oxidize a cysteine residue (C328) to enable a more specific crosslink [42]. Despite the prevalence of divalent cations in the cell, much of the vimentin is not in bundles; a recent super-resolution imaging study of cellular vimentin filaments revealed that their persistence length is as low as that of a single vimentin filament *in vitro* [45], suggesting that cellular mechanisms may exist for regulating filament bundling.

### 3.2 Cytoskeletal protein interactions controlling vimentin network assembly

Within the cell, there are more interactions responsible for vimentin network structure and formation than just the counterions. The cytoskeletal polymers of actin, microtubules, and intermediate filaments form an composite network in the cell formed by many different interacting species [46]. Vimentin, actin and microtubules interact through specific crosslinkers, motor proteins, and steric constraints that mediate the cytoskeletal structure and organization in the cell [47, 48]. F-actin and vimentin can bind to each other *in vitro* [49] and interact directly via the vimentin tail domain [50, 51]. Their interaction in mixed networks *in vitro* leads to a stiffer reconstituted network than either of the two polymers alone [52]. This stiffening is dependent on the concentration of F-actin crosslinkers; for composite networks with scarce F-actin crosslinkers, vimentin softens the network by imposing steric constraints on F-actin fluctuations needed for crosslinking [53]. Several of actin's crosslinking proteins, such as fimbrin, calponin, and filamin A, also interact with vimentin [36, 54, 55] as well as its motor protein myosin [56]. As for microtubules, they are responsible for helping stabilize vimentin during its assembly *in vivo*, with the aid of motor proteins kinesin [57], dynactin, and dynein [58]. The protein plectin plays an important role as a stabilizing crosslinker for the vimentin networks [59] as well as forming bridges between vimentin, F-actin, and microtubules [60].

### 3.3 Factors regulating vimentin filament assembly and disassembly

Reorganization of the vimentin network entails the assembly and disassembly of vimentin filaments that is needed for various cellular functions including migration and mitosis. Fully polymerized vimentin networks are very stable in cells in comparison to actin or microtubules networks [61]. Subunit exchange between soluble vimentin tetramers and the fully formed vimentin network does occur, but infrequently [62]. Vimentin networks assemble spontaneously in ionic environments *in vitro*, and after initial assembly of the unit-length filaments (ULFs) no ATP consumption is necessary for the larger hierarchical polymerization of the vimentin network *in vivo* [63]. Post-translational modifications (PTMs) of vimentin offer additional avenues for regulation [64], where the fully polymerized vimentin network may be cleaved or altered by enzymes.

One of the most common and well-studied post-translational modifications controlling cellular vimentin dynamics is phosphorylation [64, 65]. Phosphorylation is controlled by various enzymes called kinases, which bind to specific sites on the vimentin protein and promote the covalent attachment of phosphate to specific amino acids [66].

Reciprocally, dephosphorylation involves phosphatases, which remove phosphoryl groups. This regulation allows the relatively stable vimentin network to exchange with the small concentration of soluble vimentin tetramers and ULFs in the cell [67]. Phosphorylation of the vimentin N-terminus generally increases the off rate of these exchanges, thereby encouraging the disassembly of the vimentin networks, although phosphorylation at other sites can lead to increased vimentin assembly [68] depending on the mechanical state of the cell [7]. Disassembly of vimentin is an important structural requirement for mitosis of the cell [69] as well as for the formation of lamellipodia during cell extension [70]. Electrophilic species such as cyclopentenone prostaglandins (cyPCs) have been shown to disassemble vimentin networks in cells as well [71]. Another mechanism of vimentin disassembly is proteolysis; enzymes such as caspase [72] and calpain [73-75] promote vimentin IF disassembly for different cell functions by irreversibly cleaving filaments within the vimentin network. The assembly state of cellular vimentin is also highly responsive to physical stimuli. Soft substrates tend to increase the solubility of vimentin in a process that depends on phosphorylation [7], and hypoosmotic shock leads to rapid and reversible disassembly of the vimentin network [73, 74, 76] by activation of calpain [77]. Other post-translational modifications to vimentin include: glycosylation, which mediates protein-protein interactions and is important for cell migration [78]; ubiquitylation, which causes vimentin aggregation [79]; and citrullination, which facilitates vimentin disassembly [80].

## 4. Vimentin and nuclear mechanics

Vimentin intermediate filaments are closely associated with the cell nucleus and form a cage-like network that encircles the nucleus [7, 81, 82]. This cage is most evident in the early stages of cell spreading or when cells are grown on soft substrates, when nearly all the filaments are closely associated around the nucleus [7, 83]. As cells increase their cell spread area on stiff substrates, the vimentin filaments then begin to radiate out from the nucleus toward the cell periphery. Vimentin's close association with the nuclear surface is important to transmitting mechanical forces from the cell cortex to the nuclear envelope. One of the first demonstrations of this was done by applying forces at the cell surface through surface bound microbeads in endothelial cells [84]. Pulling at the cell cortex deforms the cell nucleus in the direction of pull. At small strain, actin microfilaments mediated force transfer to the nucleus; however, at large strain the actin gel can tear. In contrast, intermediate filaments effectively mediated force transfer to the nucleus under both conditions.

A common feature of vimentin-expressing mesenchymal cells is their propensity for cell migration and the routine subjection to the accompanying pushing, pulling, and frictional forces. In 2D, migration rates can be largely explained by actin assembly dynamics and adhesion to the surface, with little contribution from the resistance of the nucleus and other large organelles to movement through the ambient fluid medium. In 3D tissues (Figure 6), the physical constraints are different, and nuclear shape and rigidity limits cell passage through 2-5 µm pores, which are smaller than the unstressed nuclear diameter [85].Whereas the actin-based cytoskeleton has been considered fluid-like on the time-scale relevant to pore migration (>3 hr), intermediate filament proteins, in contrast, form the most stable cytoskeletal network [86] and therefore might function as a cytoskeletal component to facilitate confined cell migration.

Emerging work points to a role in vimentin IF in helping to pull the nucleus forward through a confining space in an extracellular matrix [87]. Vimentin is known to be indirectly connected to the nucleus via plectin and nesprin-3 [88]. In a recent study [87], the authors found a direct interaction between vimentin, nesprin-3 and actin-myosin in primary human fibroblasts, revealed by immunoprecipitation. Fluorescence images showed that vimentin and actomyosin accumulated in front of the cell nucleus. Depleting nesprin-3 led to a loss of independent movement of the nucleus, suggesting that vimentin through its interactions with nesprin-3 and actomyosin might contribute to a contractile pulling unit required for 3D cell migration.

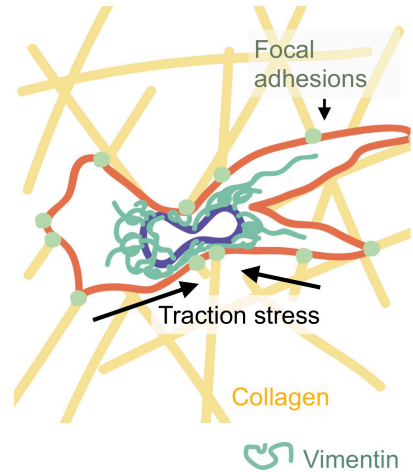

Figure 6. Schematic of a cell migrating through the extracellular matrix.

New research is also revealing that vimentin protects the nucleus from damage during migration. The extreme strains associated with confined migration can lead to structural damage to the nucleus (Figure 7). This includes nuclear blebs, accumulated double-stranded (ds) DNA breaks, and nuclear envelope rupture, which leads to unregulated mixing of nuclear and cytoplasmic materials [89, 90]. To investigate vimentin's role in facilitating confined cell motility, we developed 3D micro-fluidic channels that mimic the 3D nature of tissues [91]. Our results showed mouse embryonic fibroblast (mEF) cells lacking vimentin move much faster than control cells in our 3D system. Cells lacking vimentin exhibited higher nuclear deformations after migrating through pores of 3 µm diameter, much smaller than the effective diameter of an unstressed mEF nucleus (~10 µm). Loss of vimentin also increased the rates of nuclear damage (e.g. nuclear blebs, nuclear envelope rupture, and dsDNA breaks) associated with confined motility [81]. Loss of vimentin did not alter nuclear lamin levels [81], suggesting that the enhanced nuclear damage observed in vimentin-null cells was independent of their effect on nuclear stiffness. Taken together, the results suggest that the perinuclear vimentin cage enhances the effective stiffness of the nuclear envelope and could serve to cushion the nucleus or the cortical actin network during extreme strains associated with confined migration.

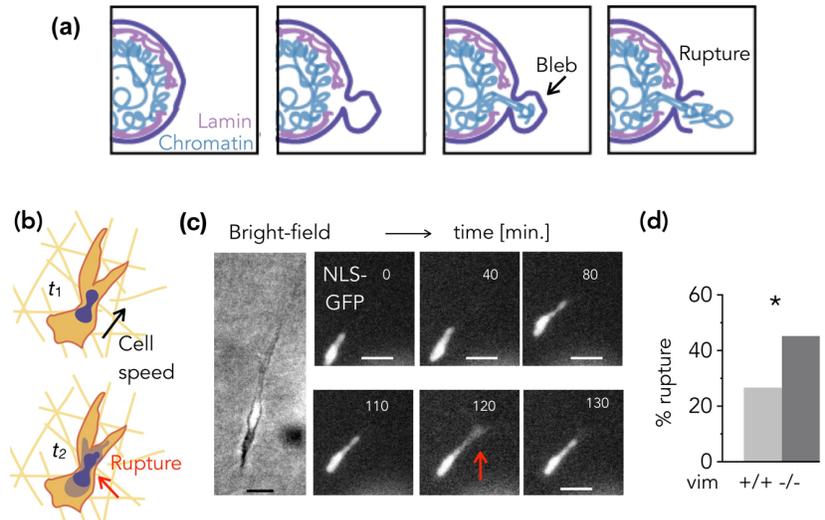

Figure 7. (a) Schematic of nuclear envelope (NE) rupture. (b) Cells embedded in 3D collagen matrices exhibit spontaneous NE rupture. (c) Loss of NLS-GFP signal from the nucleus into the cytoplasm indicates NE rupture, and (d) cells lacking VIF exhibit increased NE rupture.

While vimentin is an abundant cytoskeletal protein, its role in nuclear shape is poorly understood. On 2D surfaces, nuclear morphology is thought to be predominantly due to a balance between actomyosin contractile forces that compress the nucleus and the microtubule filaments that resist compression [92, 93], although nuclear flattening is seen in spread cells in the absence of myosin activity [94]. Recent studies have now shown that disruption or deletion of the vimentin network alters nuclear shape, reducing nuclear flattening in cells spread on 2D substrates [81, 93, 95]. These studies show that vimentin contributes to nuclear flattening, though the mechanism is not yet clear. One possible explanation is via vimentin's linkages to the nucleus [88] and the actomyosin contractile units (plectin, filamin A and other cross linkers). Because deformation of the nucleus by applied forces can directly affect gene expression [96], delineating vimentin's role in nuclear deformability is important to understanding how cells sense and respond to mechanical cues from their extracellular environment.

## 5. Vimentin in cells: Mechanics and single cell motility

### 5.1 Cell mechanics

Intermediate filaments are important for protecting cells and tissues from stress. For example, keratin is expressed in keratinocytes in the skin, but point mutations in keratin can cause skin-blistering diseases that lead these normally tough epithelial cells to break down when the skin is subject to mild everyday stretching or rubbing [97, 98]. Vimentin IFs, on the other hand, are expressed in mesenchymal cells, such as fibroblasts and endothelial cells, which are also routinely subjected to mechanical stresses. This includes pushing and pulling forces associated with cell motility but also the shear forces generated by blood and airway surface fluid flows. One of the earliest studies on the role of vimentin in the mechanical response of cells was done with endothelial cells expressing GFP-labeled vimentin and subjected to shear flow [99, 100]. Shear stress acting on the surface of the endothelium results rapidly deforms the vimentin network, indicating a role in VIFs in redistributing the intracellular forces in response to hemodynamic shear. The tension in endothelial cells is sensed and transduced by a transmembrane complex containing VE-cadherin and PECAM adhesion receptors. Tension on PECAM-1 is mediated by its association with vimentin, which is triggered by flow-stimulated mechanical stress [101]. These effects could contribute to how hemodynamic shear stress is transduced by the endothelium and its role in regulating vasoactivity [102]; notably, vimentin-null mice have been found to have impaired vasoactivity in the ability to properly remodel arteries [103].

The viscoelastic properties of vimentin filaments suggest general features that are relevant to their function in the cell. Mechanical measurements of the cell and the effects of vimentin, however, have been difficult to probe directly. This is due in part to the heterogeneous nature of the cell. Even in simple mechanical pictures, the cell is a complex bag. One could imagine the cell as a gooey droplet, covered in a stiff actin-rich cortical shell, stuffed with a dynamic cytoskeleton and a large rigid nucleus with IFs concentrated around it, deep inside the cell (Figure 8). As one might expect, mechanical measurements are thus strongly dependent on the way in which the cell is probed. The most commonly employed techniques to characterize the mechanical properties of individual cells are by atomic force microscopy (AFM), optical magnetic twisting cytometry, and active micro-rheology/optical tweezers. These studies have been performed on mesenchymal cells with a disrupted or no vimentin network. The results show the complex and varied effects of vimentin on cell stiffness, as summarized in Table 1. For example, loss of vimentin has been shown to either increase or decrease cortical stiffness depending on the thickness of the cell [104]. Often the effect of VIF on cell mechanics is not readily apparent, especially when cells are minimally deformed or probed only in the actin-rich cortex [105]. However, more studies are emerging showing that VIF do indeed increase cell elastic behavior, particularly for large [106] and repeated deformations [105] and in the perinuclear [91] and cytoplasmic region of the cell [107].

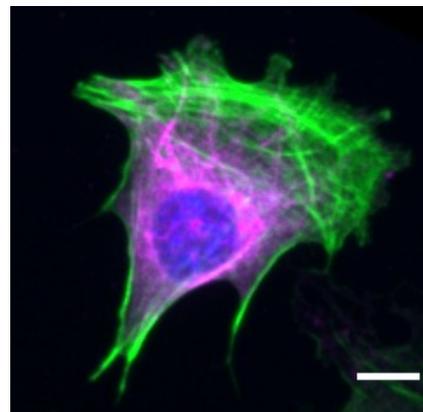

Figure 8. Immunofluorescence image of vimentin inside the cell. Image of a fixed mouse embryo fibroblast, labeled in green for F-actin, magenta for vimentin, and blue for DNA. Scale bar, 10 μm.

These studies paint a more complete picture of vimentin IF, setting them apart from both actin filaments and microtubules in preserving the structural integrity of the cell and nucleus. Vimentin IF are sufficiently soft to allow moderate cellular deformations without making the cytoplasm too rigid or brittle. Yet, under strong stresses that can rupture the actin network, the strain stiffening of vimentin prevents excessive deformation. Recent experiments demonstrate how vimentin is important for maintaining the basic integrity of the cell. These effects are most evident when cells are subjected to exceedingly large deformations as expected by the strain-stiffening mechanical properties of VIFs. Two examples of such experiments include single cell migration through confining spaces as well as deformation of 3D hydrogel matrixes embedded with cells. For example, VIFs have little effect on reducing cell speed and nuclear damage when cells move through relatively large channels, on the order of the cell size or larger (> 10 μm), but kick in for smaller, more confining spaces (3 to 10 μm), where cellular strains are larger [81, 91]. Extreme extensions [108] and compressions [81] of hydrogels embedded with cells show that VIFs support cell viability, significantly reducing the amount of whole cell damage and lysis that occurs at large strains.

| Method | Main Results | Reference |
| --- | --- | --- |
| Rotational twisting cytometer | Loss of vimentin decreases cortical rigidity | Wang & Ingber, 1994 [109] |
| Rotational twisting cytometer | Loss of vimentin decreases cortical rigidity only at large strains | Wang & Stamenovic, 2000 [106] |
| Atomic force microscopy | Vimentin's effect on cell stiffness only apparent for maximally-spread cells or for cells under repeated deformations | Mendez, Restle, Janmey 2014 [105] |
| Rotational twisting cytometer, active micro-rheology | VIF contribute little to cortical stiffness but doubles cytoplasmic shear modulus | Guo, Ehrilicher, Mahammad, et al. 2013 [107] |
| Atomic force microscopy | Deletion of vimentin decreases cortical stiffness by ~10% in MDA231 breast cancer cells | Messica et al. 2017 [110] |
| Rotational twisting cytometer Atomic force microscopy | Loss of vimentin decreased cell stiffness as measured by sharp-tip and round-tip AFM probes; yet, loss of vimentin increased cortex stiffness reported by twisting cytometry, attributed to reduced cell thickness | Vahabikashi, Park, Perkumas, et al 2019 [104] |
| Optical tweezers | VIF increases cytoplasmic stiffness and toughness | Hu, Li, Hao, et al. 2019 [108] |
| Atomic force microscopy | VIF increases cell stiffness only in the perinuclear region of the cell | Patteson, Pogoda, Byfield et al. 2019 [91] |
| Atomic force microscopy | Re-expression of VIF in vim -/- mEFs rescues cell stiffness | Patteson, Vahabikashi, Pogoda, et al. 2019 [81] |

## 5.2 Cell motility

Mesenchymal cell migration depends on coordinated regulation between the cytoskeletal network and focal adhesions complexes, sites of contact between the cell and their extracellular environment. Intermediate filaments have been considered as passive in the context of cell migration, in contrast to actin polymerization, which actively drives cellular protrusions. However, recent work is revealing a range of non-mechanical functions for vimentin that suggest a more active role in interacting and regulating cytoskeletal dynamics that drive cell motility by interacting with multiple signaling pathways [111-113]. Here, we will review a number of these studies but will refer the readers to a recent review for more details [47].

Most studies on the effects of VIF on cell motility have been conducted on unstructured 2D glass or tissue culture plastic substrates. One of the first studies was in 1998 using primary mouse embryo fibroblasts derived from wild-type and vimentin-null mice and showed reduced cell motility in both wound healing assays and directional migration towards a chemo-attractant [114]. The investigators found irregular stress fiber organization and large clusters of focal adhesion structures in vimentin-null cells, 'as if they failed to remodel into smaller, discrete and separated adhesion complexes [114].'

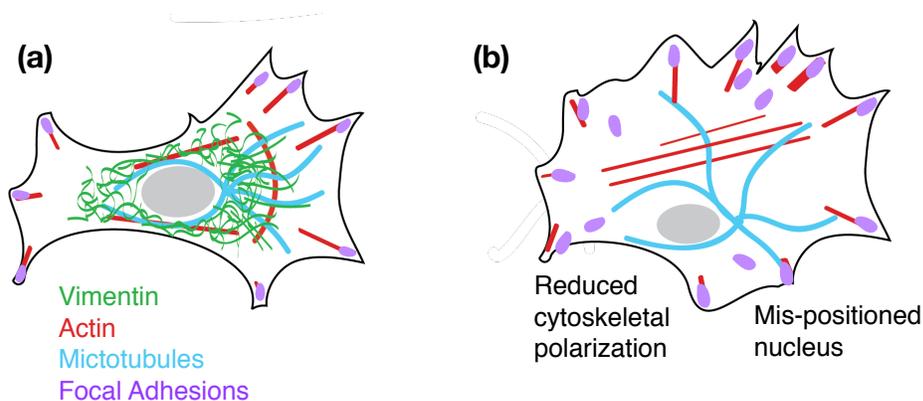

Figure 9. Schematic representation of cytoskeletal organization in cells with (a) and without vimentin IFs (b).

Why does vimentin increase 2D cell motility? One might suspect that vimentin assists in increased actin polymerization or actin stress fiber formation; however, the opposite seems to be true. Loss of vimentin stimulates actin-based lamellipodia formation [70] and increases stress fiber formation through GEF-H1 and Rho-A [111]. Instead the answer seems to lie more in vimentin's role in promoting a polarized cell shape that is characteristic of mesenchymal cells [70, 112, 113, 115] (Figure 9). A polarized cell shape is a critical determinant in establishing coordinated migration. Loss of vimentin tends to increase lamellipodia formation continuously around the circumference of the cell, pulling the cell in all directions while translocating it in none [70]. In the presence of vimentin fibers, actomyosin-generated forces are partially absorbed by the vimentin network and are redirected to adhesions [113]. As a result, vimentin contributes to the alignment of traction stresses that a cell applies to its substrate, which permit single cell migration. On unstructured flat substrates, microtubules are an important player in establishing front-rear cell polarity [116]. Vimentin filaments form a coextensive network with microtubules [117] and are long-lived compared to microtubules. Recent experiments have shown that vimentin filaments act as a template for future microtubule growth along previous microtubule tracks, preserving the positioning of microtubule networks and the polarity of the cell [112]. These results might also be relevant to the finding that loss of vimentin impaired motility of breast cancer cells in 2D only in dense cultures but had no effect in sparse culture [110].

The interactions of cells with the extracellular matrix are facilitated through the activation of focal adhesion (FA) complexes. Studies show that vimentin is connected to FAs and that this connection is required for the transfer of mechanical forces from the cell surface to the cell interior [118]. Fluorescence recovery after photobleaching (FRAP) studies of GFP-labeled paxillin, a focal adhesion protein, showed that vimentin increases the rate of focal adhesion turnover by up to 400%, indicating that FA are much more dynamic in vimentin-expressing cells [115]. Collagen I is the most abundant extracellular matrix protein, and recent studies show that cell adhesion to collagen via β1 integrins is dependent on the actin cross-linking protein filamin A and vimentin [68, 119, 120]. The effects of vimentin on focal adhesion assembly remains largely unclear, in part because vimentin does not generally concentrate sufficiently at these sites to be evident by immunofluorescence imaging the way it does at desmosomes or hemidesmosomes that engage different transmembrane receptors [121, 122]. An exception to this rule is the finding that in the lung cancer cell line A549, complexes containing β4 integrin cluster along vimentin at the cell periphery. The binding is not direct but is mediated by the crosslinking protein plectin [123]. Although vimentin deficient A549 cells still contain β4 integrin adhesion sites, these cells move more slowly, and the important actin regulator Rac1 fails to localize to β4 integrin [124]. These results are consistent with the hypothesis that vimentin alters actin assembly in part by its effect on the small GTPases such as RhoA and Rac1 that control actin assembly. Identifying vimentin's role in focal adhesion activation will be important to interpreting how cells establish the strength and coordination of traction stresses that enable cell migration.

Tissues provide cells with a 3D architecture with distinct physical challenges that fundamentally alter the mechanisms that cells use to generate and transmit forces in order to move [125, 126]. The effects of disrupting specific proteins on 2D motility can differ with their effects in 3D [127, 128]. For example, regulation of 2D cell motility by focal adhesion proteins is not predictive of regulation of 3D cell motility in matrix [129].

Along with this trend, we found that loss of vimentin increases cell motility in 3D channels, even though it has the opposite effect in 2D [81, 91]. The essential function of vimentin in motility of mesenchymal cells in 3D has been tested with three different systems: microfluidic channels, 3D collagen matrices, and thin curved surfaces to mimic both the complex geometry and the positive and negative curvatures that these cells encounter *in vivo*. All three methods show similar effects of vimentin that are absent from conventional studies in 2D. The numerous functions vimentin plays in 2D motility are likely still present in these 3D systems, but one possible explanation for the increased motility in cells lacking vimentin is that the confining channels and 3D spaces require large cellular strains, which are resisted by the strain-stiffening properties of the vimentin networks. Another possibility is that in structured environments, cells are presented with an external geometric cue that aligns and polarizes the cell. Now in a polarized cell state, actin-based lamellipodia (which is stronger in vimentin-null cells) takes over and increases cell migration.

Recent studies have investigated the role of vimentin in amoeboid confined motility [130, 131]. Amoeboid migration is a rapid mode of motility that is characterized by low adhesion and blebbing [132, 133]. In human cancer cells, such as melanoma A375-M2 cells and MDA-MB-231 breast cancer cells, RNAi depletion of vimentin increased amoeboid motility by 50% compared to control cells [131]. In contrast, vimentin was shown to have the opposite effect in dendritic cells derived from wild-type and vimentin-null mice [130]. Dendritic cells and their capacity to migrate through small spaces are necessary for the immune response [134]. It was shown that vimentin increases amoeboid migration in confining environments and that loss of vimentin reduced actin mobility in the cell cortex. Vimentin-deficiency also led to DNA double strand breaks in the compressed dendritic cells.

The varied effects of vimentin on cell motility in different cell lines and experimental conditions (e.g. 2D vs 3D) might seem counter-intuitive. On the other hand, vimentin contributes to multiple cytoskeletal remodeling pathways and the structure of vimentin can be altered by subunit exchange, cleavage into different sizes, re-annealing, post-translational modifications and dozens of interacting proteins [135]. Vimentin has over 53 phosphorylation sites, many of which are targets for kinases involved in the induction of lamellipodia and cell motility [70]. Phosphorylation of vimentin sites typically leads to disassembly of filaments as well as changes in bundled assembly status [67, 136]. The spatial regulation of specific phosphorylation sites vimentin sites has been implicated in cell motility. For example, phosphorylation of vimentin at Ser-38 by Rac1's target PAK, p21-activated kinase, leads to vimentin network disassembly and local induction of lamellipodia formation [70]. In addition, recent experiments suggest that the highly migratory ability of transformed metastasizing tumor cells depends not on vimentin expression levels but rather on spatially-controlled phosphorylation states of vimentin at serine 71, which is phosphorylated by RhoA's target ROCK [137] (Figure 10). The diversity of vimentin's biological functions confers advantageous benefits to the cell, allowing it to readily adjust to different experimental conditions and selective pressures.

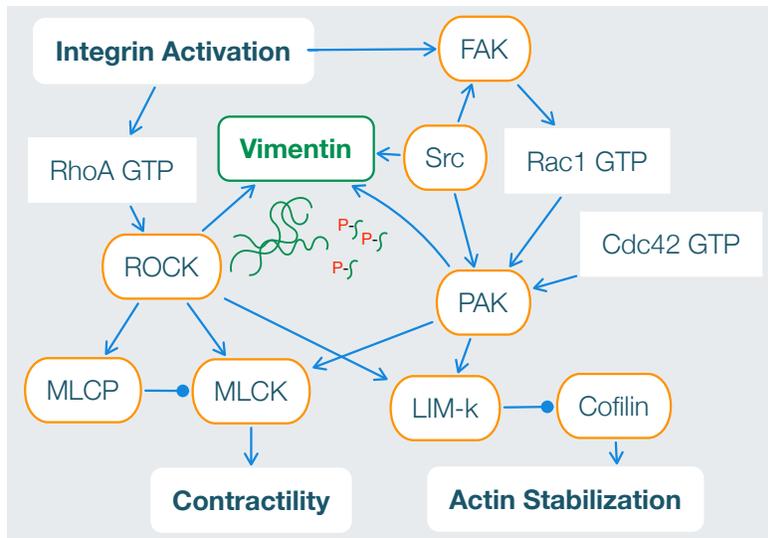

Figure 10. RhoA and Rac1 signaling pathways involved vimentin network assembly. Schematic adopted from [7].

### 5.3 Vimentin in cell division

At the onset of cytokinesis, the process by which the mother cell divides into two daughter cells after the nucleus has duplicated, the vimentin network is dissembled by the activity of multiple kinases that phosphorylate vimentin at its N-terminal domain [138-140]. Mitosis, the duplication of nuclei that precedes cytokinesis, also activates kinases that disassemble vimentin [141]. However, rather than simply removing the viscoelastic impediment to nuclear and cell division, the relocation of vimentin during the cell cycle actively contributes to this process. Cytokinesis is an acto-myosin driven process, and a recent proteomic analysis showed that vimentin, presumably released from the disassembled perinuclear cage, is concentrated at the actin-rich cortex in a plectin-dependent manner [142]. Moreover, the unstructured C-terminus of vimentin is essential for this localization and normal mitotic progression. If the vimentin C-terminus is deleted or the actin cortex is disrupted, vimentin forms bundles that associate with chromosomes and disrupt mitosis [143]. Vimentin also appears to be important for establishing asymmetry in mitosis, and by asymmetric distribution vimentin can concentrate misfolded proteins destined for degradation in one of the two daughter cells, a process that is related to the cellular response to aging [144]. These results emphasize the active role of vimentin in specialized processes related to cell division in animal cells.

## 6. Conclusion

The filaments and filament assemblies made by the intermediate filament protein vimentin have multiple cellular functions, many of which depend on the unique physical properties of these biopolymers.  Vimentin is much softer than the other cytoskeletal filaments, F-actin and microtubules, but it is also the most dramatically strain-stiffening and the most stable against damage from mechanical stress.  This robustness of vimentin filaments is a consequence of the numerous dynamic bonds that form between the alpha-helical coiled coil subunits within the filaments. These interactions are affected by the strongly anionic surface charge of the filament.  As a result, electrostatically-dominated crosslinks and the multiple bonds within the filaments can resist significant forces, but also move and reform at high stresses to allow the filaments to dissipate mechanical energy without breaking.  Not all cellular functions of vimentin depend on its physical properties, but this combination of unique polymer physics and multiple interactions with other cellular structures, including the nucleus, lead to cellular effects that are essential for the normal functioning of animal cells in a mechanically stressful environment.